\documentclass[aps,pre,showpacs,onecolumn,superscriptaddress]{revtex4}

\usepackage{amsmath}
\usepackage[final]{listings}
\usepackage{color}
\usepackage{hypernat}
\usepackage{graphicx}
\definecolor{mylinkcolor}{rgb}{0,0,0.7}
\definecolor{mycitecolor}{rgb}{0.0,0.6,0.0}
\usepackage[citecolor=mycitecolor,colorlinks=true,linkcolor=mylinkcolor,plainpages=false]{hyperref}

\newcommand{\rem}[1]{}

\begin{document}
    
\title{Odeint -- Solving ordinary differential equations in C++}


\author{Karsten Ahnert}
\author{Mario Mulansky}
\affiliation{Department of Physics and Astronomy, University of Potsdam}

\maketitle

\section{Introduction}

Ordinary differential equations (ODEs) play a crucial role in many scientific disciplines. For example the Newtonian and Hamiltonian mechanics are completely formulated in terms of ODEs. Other important applications can be found in biology (population dynamics or neuroscience), statistical physics and  molecular dynamics or in nonlinear sciences~\cite{Ott-book-02}. Furthermore, ODEs are used in numerical simulations to solve partial differential equations (PDEs), for example by discretizing the spatial coordinates.

An analytic solution of an ODE can only be found in very rare cases such that numerical methods have to be employed. Solving ODEs numerically~\cite{HairerSolvingODEI,HairerSolvingODEII} has a long tradition and has become popular by the rise of computers and its spreading in form of micro computers which are now present in our daily live. The most famous solvers for ODEs are surely the well-known explicit Runge-Kutta solvers which are easy to implement and can easily be applied to a wide range of problems. They come with step-size control and some algorithms also possess dense output functionality. Another class of solvers are implicit solvers which are important for stiff problems, hence ODEs with two or more different scales of the independent variable.

In mathematical terms, an ordinary differential equation is defined as
\begin{equation}
 \dot{\vec{x}} = \vec{f}(\vec{x} , t )
\,\,\text{.}
\end{equation}
Here and in the following the time $t$ is used as the independent variable. The state of the ODE is $\vec{x}$ which is a vector field and $\dot{\vec{x}}$ denotes its time derivative. An initial value problem (IVP) of an ODE is to find a solution given an initial value $\vec{x}_0(t_0)$. Nearly all methods for solving ODEs work iteratively, that is they start with $\vec{x}_0(t_0)$ and iteratively create a sequence $\vec{x}(t_i)$ where every $\vec{x}(t_i)$ is obtained from previously calculated values of $\vec{x}$.

In this paper we introduce odeint~\cite{Odeint2011} -- a C++ library for solving the IVP of ODEs. Of course, there exist many of such libraries for different languages. Particular examples are the GNU Scientific Library (gsl)~\cite{GSL2011} or the routines from the Numerical Recipes (NR)~\cite{NR}. These two libraries are written in C and C++ and are widely used in the scientific community. Other popular examples are Apache.Math~\cite{ApacheMath2011} for Java, the ode* functions MATLAB~\cite{Matlab2011} or scipy.integrate.odeint~\cite{Scipy.odeint2011} for Python.

\section{Requirements}

The main goal of odeint is to provide a modern and fast C++ library for solving the initial value problem (IVP) of ODEs. Furthermore, emphasis is put on the following points:

\textbf{Container independence} -- The state type of the ODE should be parametrized by the user of odeint. It should be possible to use odeint with the most common types in C++ such as \texttt{vector< double >} or \texttt{array< double , N >}. It must also be possible to work in the complex domain simply by using \texttt{array< complex< double > , N >} as representation for the state. This is already a major improvement over most of the existing libraries. Furthermore, it must be possible to use odeint with exotic state types like matrices, complex networks, and vectors or arrays living on modern GPUs.

\textbf{Operation independence} -- It must be possible to change the way numerical operations are performed. This way one can use odeint with SIMD (Single instruction multiple data) operations and arbitrary precision types.

\textbf{High performance} -- Odeint should be fast and its performance must be at least comparable to standard software on ODEs like gsl and NR.

\textbf{Generality} -- The design of odeint must be generic, such that it is possible to implement arbitrary solvers within the framework and its interfaces. It must support the classical Runge-Kutta steppers, implicit methods and solvers for stiff systems, as well as symplectic solvers and multistep methods. Step size control must be implement and it should use dense output functionality if the solver under consideration supports it.

\section{Library structure and design}

Odeint is an open source library. It is available via subversion or by direct download~\cite{Odeint2011}. Odeint lives within the boost ecosystem~\cite{Boost2011} -- a collection of state-of-the-art C++ libraries. The boost libraries are well know within the C++ community for their high quality. Several C++ standard libraries have been implemented and released here. At the moment odeint is under development, therefore it is not an official boost library. It is planned to bring odeint to a level that it can be accepted as a full boost library.

The design of odeint is based on generic programming and functional programming using the advantages of the C++ template system~\cite{ModernC++,C++Templates}. Generic programming allows one to use static polymorphism also known as compile-time polymorphism to create the basic parts of the library. This has the advantage that all parts are known during compilation and the C++ compiler can use mighty optimization techniques to build fast run-time machine code. Run-time polymorphism is not sufficient here since it always results in direct function calls which can not be optimized.

\begin{figure}
\begin{center}\includegraphics[draft=false,width=1.0\textwidth]{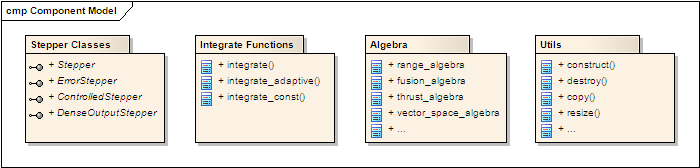}\end{center}
\caption{\label{fig:components}Brief overview over the structure of odeint.}
\end{figure}

Odeint is a header-only C++-library. So, all one has to do to use odeint is to include the appropriate headers. The overall structure of odeint is shown in Fig.~\ref{fig:components}. It consists of four basic parts: one part defines the steppers, a second one defines integrate functions which use the steppers to compute a numerical solution of the ODE. In the third part different algebras are introduced which are responsible for performing the basic operations in the most common steppers while a fourth one defines utility and helper functions, for example functions for resizing, copying, etc.

In classical object-oriented languages one uses interfaces or abstract classes to define the methods and functionality a special class has to fulfill~\cite{GoF}. In generic programming this is not possible. Here, one defines concepts which are \emph{not} written within the source code but which are defined elsewhere, for example in the documentation. A concept is a convention how a class can be used, e.g. which methods it must provide and which types it defines. In odeint four basic stepper concepts are defined.

The most general concept is the \texttt{Stepper} concept which corresponds to the basic interface one expects on a solver for ODEs. Any stepper fulfilling this concept has to have a method \texttt{do\_step( ode , x , t , dt )} which performs one iteration of the ODE defined by \texttt{ode} with the current state \texttt{x} at the time \texttt{t} with a step size \texttt{dt}. This method transforms the state of the ODE in-place, meaning that the state of the ODE is updated inside the same container. The concept also defines an out-of-place \texttt{do\_step} method. Furthermore, it must define the types to represent the state, the time, the derivative of the state, the basic value type and a function returning the order of the stepper. The ODE enters the stepper as a template parameter of \texttt{do\_step}. It can be a function pointer or a functor, hence a class with a public \texttt{operator()}. Furthermore, the function signature should be in most cases \texttt{ode( x , dxdt , t )}.

Besides the stepper concept an \texttt{ErrorStepper} concept is available which only differs from the stepper concept by the \texttt{do\_step} method. Here, this method also fills a state type containing the error made during one step \texttt{do\_step( ode , x , t , dt , xerr )}. This error might be used by an appropriate step size controller. For such controllers an own concept exist -- the \texttt{ControlledStepper} concept. It defines a method \texttt{try\_step( ode , x , t , dt )} which will try to perform a controlled step. If this trial has been successful \texttt{t} is increased, \texttt{dt} is adapted, and the state \texttt{x} is set to its new value. Furthermore, this method will return an \texttt{enum} indicating whether the current step has been accepted and the step size is unchanged or increased or if the step has been rejected. In odeint several controlled steppers exist, for example one which has an error stepper as a parameter and several specialized controlled steppers for specific steppers. Finally, the \texttt{DenseOutputStepper} concept defines steppers with dense output functionality. Here, the \texttt{do\_step}-method has the signature \texttt{do\_step( ode )}, hence a dense output stepper controls the state and the step size internally.

\begin{table}
  \caption{\label{tab:stepper_overview}Overview over the steppers implemented in odeint.}
  {\centering
    \begin{tabular}{llcl}
      \hline
      Method & Class name & Stepper concept & Notes \\
      \hline
      Explicit euler & \texttt{explicit\_euler} & S & \\
      Runge-Kutta 4 & \texttt{explicit\_rk4} & S & \\
      Runge-Kutta Cash-Karp & \texttt{explicit\_error\_rk54\_ck} & SE & \\
      Dormand-Prince 5 & \texttt{explicit\_error\_dopri5} & SE & Can be used with \\
          & & & dense output \\
      Implicit Euler & \texttt{implicit\_euler} & S & \\
      Rosenbrock 4 & \texttt{rosenbrock4} & SE & Provides a separate controller \\
          & & & and dense output \\
      Symplectic Euler & \texttt{symplectic\_euler} & S & \\
      \hline 
      Default controller & \texttt{controlled\_error\_stepper} & C & Works with explicit error steppers\\
      Default dense output & \texttt{explicit\_dense\_output} & D & Works with Dormand-Prince 5 \\
      \hline
    \end{tabular}
  }
\end{table}

Some steppers fulfill more than one of the above concepts. For further specialization several sub concepts have been defined which provide nicer interfaces to the stepper. For example the classical explicit steppers and the explicit FSAL-steppers (first same as last) have their own concepts. At the moment some standard steppers have been implemented, see Table~\ref{tab:stepper_overview}. For future versions of odeint it is planned to implement several other schemes.

The steppers can be used inside the integrate functions which allow an easy generation of the numerical solution of an ODE; \texttt{integrate\_const} generates a solution with constant step size and \texttt{integrate\_adaptive} iterates the solution with step size control. All integrate functions take full advantage of the specific stepper type. For example, if a dense-output stepper is used with \texttt{integrate\_const} it performs the largest possible steps and evaluates the solution with the help of the dense output functionality.

To access the state of the current numerical solution of the ODE every integrate function accepts an additional argument -- an observer. This observer has to be a function pointer or a functor and it can be used to write the state of the ODE or to do some statistical analysis. An example how the integrate functions and the observers play together is shown in Listing~\ref{lst:lorenz_integrate}. It is also possible to compose the observer from functional programming libraries, like Boost.Lambda.

\begin{lstlisting}[float=t,caption=Examplatory usage of odeint.,label=lst:lorenz_integrate]
typedef std::tr1::array< double , 3 > state_type;

void lorenz( const state_type &x , state_type &dxdt , double t ) {
    dxdt[0] = 10.0 * ( x[1] - x[0] );
    dxdt[1] = 28.0 * x[0] - x[1] - x[0] * x[2];
    dxdt[2] = - 8.0 / 3.0 * x[2] + x[0] * x[1];
}

inline void write_state ( const state_type &x , double t ) {
    std::cout << t;
    for( size_t i=0 ; i<x.size() ; ++i ) std::cout << "\t" << x[i];
    std::cout << "\n";
}


explicit_rk4< state_type > rk4;
state_type x = {{ 10.0 , 10.0 , 10.0 }};
for( size_t i=0 ; i<1000 ; ++i )
    rk4.do_step( lorenz , x , 0.0 , 0.01 );

explicit_error_dopri5< state_type > dopri5;
integrate_const( dopri5 , lorenz , x , 0.0 , 1000.0 , 1.0 , write_state );
\end{lstlisting}

As stated above one of the main requirements of odeint is its container independence. For this reason an algebra concept has been introduced which is used by most of the steppers (but not by all). It defines how basic operations on a state type are performed. Steppers taking advantage of the algebra concept are parametrized by the algebra. This gives very interesting possibilities for high performance computing using modern GPUs and is one of the major advantages of odeint. For example, it is possible to perform parameter studies of a given ODE on a GPU where a whole ensemble of ODEs with different parameters is iterated in parallel. Or it is possible to study large lattices of coupled ODEs or discretized PDEs on a GPU. Examples how one can use odeint with GPUs and CUDA are shown in the documentation of odeint.

Other examples where one can use the advantages of the algebras within odeint are exotic state types. For example, it is possible to create algebras which work with the containers from the GSL. Another interesting use case are ODEs defined on regular $n$-dimensional lattices where the underlying state type is a container with $n$-indices. Again, in this case one has to specialize the algebra and one can use all routines from odeint. ODEs on complex networks and graphs are also possible to solve with odeint as far as an appropriate algebra has been implemented.

\section{Conclusion}

In this article odeint -- a high level C++ library for solving ordinary differential equations has been introduced. Its main advantages over existing solutions is its performance and its container independence making this library feasible for many use cases ranging from educational purposes to high-performance computing. During the development of odeint many advanced programming techniques like template meta programming, expression templates and functional programming have been used. For example, a new technique for implementing the explicit Runge-Kutta steppers has been developed as well as a Taylor stepper of arbitrary order, where the first $k$ derivatives of the ODE are evaluated using expression templates and auto differentiation. It is planned to continuously expand odeint with new implementations of existing steppers. Furthermore, odeint may serve as a playground for research on numerical solution of ordinary differential equations.


\end{document}